%% file: main.tex
\documentclass[aps,pra,reprint,superscriptaddress]{revtex4-2}

\usepackage[english]{babel}
\usepackage{graphics}
\usepackage{graphicx}
\usepackage{amsfonts}
\usepackage{amsmath}
\usepackage{amssymb}
\usepackage{color}
\usepackage{physics}
\usepackage{tikz}
\usepackage{listings}
\usepackage[separate-uncertainty,per-mode=fraction]{siunitx}
\DeclareSIUnit\gauss{G}

\usepackage{threeparttable}
\usepackage{booktabs} 

\usepackage{todonotes}

\definecolor{darkblue}{rgb}{0.0,0.0,0.4}
\definecolor{darkgreen}{rgb}{0.0,0.4,0.0}
\definecolor{darkred}{rgb}{0.6,0.0,0.0}
\usepackage[bookmarksopen]{hyperref}
\hypersetup{colorlinks,linkcolor=darkblue,citecolor=darkblue,urlcolor=darkblue}

\newcolumntype{M}[1]{>{\centering\arraybackslash}m{#1}}

\newcommand{\etal}{\textit{et al.}}

\newcommand*{\Ham}{\mathcal{H}}

\renewcommand{\vec}[1]{\boldsymbol{#1}}

\newcommand{\Op}[1]{\boldsymbol{\hat{#1}}} 
\newcommand{\mol}[2]{${}^{#1}${#2}}
\newcommand{\BaFb}{{${}^{138}$BaF}} 
\newcommand{\gs}{{$X^{2}\Sigma^{+}$}}
\newcommand{\exs}{{$A{}^{2}\Pi_{1/2}$}}

\newcommand{\exsb}{{$B{}^{2}\Sigma$}}
\newcommand{\Nu}{n^{\text{e}}}
\newcommand{\Nl}{n^{\text{g}}}
\newcommand{\densM}[2]{\ev{ \ket{#1}\bra{#2} }}
\newcommand{\melS}[3]{\ _\text{s} \!\mel{#1}{#2}{#3}_\text{s}}
\newcommand{\Py}[1]{\texorpdfstring{\lstinline[language=Python,basicstyle=\ttfamily]{#1}}{#1}}
\newcommand{\transition}[2]{$#1 \rightarrow #2$}
\newcommand{\transitionXA}[2]{\transition{\text{\gs}(#1)}{\text{\exs}(#2)}}

\newcommand{\cooltrans}{\transition{\text{\gs}(\nu=0)}{\text{\exs}(\nu'=0)}}

\newcommand{\Wjt}[6]{ \begin{pmatrix}
		#1 & #2 & #3 \\
		#4 & #5 & #6
\end{pmatrix}}

\begin{document}

\title{Numerical modeling of laser cooling in molecules:\\ From simple diatomics to polyatomics and radioactive species
}

\author{Felix Kogel}
\affiliation{5. Physikalisches  Institut  and  Center  for  Integrated  Quantum  Science  and  Technology,Universit\"at  Stuttgart,  Pfaffenwaldring  57,  70569  Stuttgart,  Germany}

\author{Tatsam Garg} 
\affiliation{Vienna Center for Quantum Science and Technology, Atominstitut, TU Wien,  Stadionallee 2,  1020 Vienna,  Austria}

\author{Phillip Gro\ss}
\affiliation{Vienna Center for Quantum Science and Technology, Atominstitut, TU Wien,  Stadionallee 2,  1020 Vienna,  Austria}

\author{Lukas Leczek}
\affiliation{Vienna Center for Quantum Science and Technology, Atominstitut, TU Wien,  Stadionallee 2,  1020 Vienna,  Austria}

\author{Marian Rockenh\"auser}
\affiliation{5. Physikalisches  Institut  and  Center  for  Integrated  Quantum  Science  and  Technology,Universit\"at  Stuttgart,  Pfaffenwaldring  57,  70569  Stuttgart,  Germany}
\affiliation{Vienna Center for Quantum Science and Technology, Atominstitut, TU Wien,  Stadionallee 2,  1020 Vienna,  Austria}

\author{Neil Shah}
\affiliation{Vienna Center for Quantum Science and Technology, Atominstitut, TU Wien,  Stadionallee 2,  1020 Vienna,  Austria}

\author{Jakob Wei\ss}
\affiliation{Vienna Center for Quantum Science and Technology, Atominstitut, TU Wien,  Stadionallee 2,  1020 Vienna,  Austria}

\author{Andreas Schindewolf}
\affiliation{Vienna Center for Quantum Science and Technology, Atominstitut, TU Wien,  Stadionallee 2,  1020 Vienna,  Austria}

\author{Tim Langen}
\email{tim.langen@tuwien.ac.at}

\affiliation{5. Physikalisches  Institut  and  Center  for  Integrated  Quantum  Science  and  Technology,Universit\"at  Stuttgart,  Pfaffenwaldring  57,  70569  Stuttgart,  Germany}
\affiliation{Vienna Center for Quantum Science and Technology, Atominstitut, TU Wien,  Stadionallee 2,  1020 Vienna,  Austria}

\begin{abstract}
Optical Bloch equations and rate equations serve as powerful tools to model light-matter interactions from textbook-like two-level atoms to the complex internal dynamics of molecules. A particular challenge in this context is posed by molecular laser cooling, where many dozens or hundreds of levels need to be taken into account for a comprehensive modeling. Here, we present \Py{MoleCool}, a numerically efficient Python toolbox to implement and solve the corresponding differential equation systems. We illustrate both the capabilities of the toolbox and some of the intricacies of molecular laser cooling by educational examples, which range from simple Rabi oscillations to spontaneous and coherent cooling schemes for various currently studied or considered molecular species. This includes, in particular, a comprehensive modeling of laser cooling dynamics with full hyperfine structure resolution in radioactive radium monofluoride (RaF), as well as studies of other complex species such as barium monofluoride (BaF) and ytterbium monohydroxide (YbOH).

\end{abstract}

\maketitle 


\section{Introduction}
For many experiments in which atoms or molecules interact with laser light fields and external static fields, it is an important goal to make predictions about the particles' motion, internal state dynamics or ensemble temperature given a certain laser configuration, initial spatial distribution or state~\cite{CohenTannoudji}. On a fundamental level, such predictions shed light on the underlying mechanisms that enable efficient manipulation or cooling of the particles.  On a more practical level, these predictions play a crucial role in guiding the design and development of novel experimental setups. 

The more complex the internal structure of the particles under consideration, the more complex such simulations can become, placing high demands on the corresponding numerical implementations. This particularly applies to the laser cooling of molecules, which has recently seen intense progress~\cite{Fitch2021, McCarron2018a,Langen2023} and which can involve many dozens of internal states~\cite{Mitra2020,Kogel2025,PadillaCastillo2025}.

Depending on the desired level of detail, the required calculations can be realized by using either rate equations or optical Bloch equations (OBEs). The simpler rate equations neglect coherent effects, while the OBEs describe the full density matrix and thus allow for a more comprehensive modeling of the light-matter interactions. However, compared to the rate equations, the OBEs are numerically much more costly to solve. Previous work has used both approaches in a variety of settings, with different levels of abstraction and using a wide range of numerical techniques~\cite{Kogel2021,Grasdijk,QiSun,Riesch2021,ADM,Johansson2012,Eckel2022,Miller2023,Downes2023}. 

Here, we introduce \Py{MoleCool}, a Python toolbox to implement and solve multi-level rate equations and optical Bloch equations. Our toolbox is specifically optimized for the low-level modeling of laser cooling in complex molecules, where an efficient numerical implementation is essential to capture a large number of internal states. The toolbox includes auxiliary modules that provide calculations of the energy levels and matrix elements for several molecules. Additional data for further molecules, for example, from separate Python packages~\cite{Blackmore2023,Humphreys2025} or \Py{PGOPHER}~\cite{Pgother}, can be imported. Crucially, \Py{MoleCool} provides direct access to all individual quantum states and transition pathways, including branching ratios, thus enabling detailed modeling of population dynamics and optical forces at the level of the entire internal structure. In this way, it complements existing, more abstract approaches such as \Py{PyLCP}~\cite{Eckel2022}. 

In this work, we introduce our toolbox and highlight its features through a number of instructive examples, ranging from the textbook-like analysis of two- and three-level systems to the intricate and sometimes counter-intuitive dynamics in multi-level systems. We then address some state-of-the-art problems in the manipulation and cooling of molecules. All corresponding codes and extensive documentation are available in an online repository~\cite{GithubLangenGroup} and can be easily extended and adapted to other molecular species and experimental situations. 

\section{Multi-level systems}
We start by briefly summarizing the two approaches to model the dynamics in multi-level atomic and molecular systems---rate equations, and optical Bloch equations. 

\subsection{Rate equations}
In our work, we consider multi-level systems consisting of a set of multiple electronic ground and excited states containing a total number of $N_\text{g}$ and $N_\text{e}$ sublevels, respectively. The sublevels are characterized by their populations $\Nl_l$ and $\Nu_u$, and the angular frequencies $\omega_{l,u}$ of the transitions that connect them. Here,  the indices $u$ and $l$ are used to label the sublevels arising from rotation, hyperfine and other splittings. We further consider a set of laser beams that can drive all transitions. These beams are characterized by an index $p$, angular frequencies $\omega_p$, wavevectors $\vec{k}_p$ and normalized polarizations $\vec{\epsilon}_p$. With this, the rate equations are given in the general form~\cite{Tarbutt2015,Fitch2021},
\begin{subequations}\label{eq:rateeqs}
\begin{align}
	\dv{t} \Nl_l &= \sum_{u,p} R_{l,u,p} \qty(\Nu_u - \Nl_l) + \sum_u \Gamma_u r_{l,u} \Nu_u ,\\
	\dv{t} \Nu_u &= \sum_{l,p} R_{l,u,p} \qty(\Nl_l - \Nu_u) - \Gamma_u \Nu_u .\label{eq:rateeqs_l}
\end{align}
\end{subequations}
Here, $R_{l,u,p}$ denotes the excitation rate and $r_{l,u}$ the excited state branching ratios, which are both discussed in more detail below. $\Gamma_u$ is the decay rate, which is equal for all sublevels belonging to the same electronic excited state. The first terms on the right-hand side thus describe absorption and stimulated emission of photons, while the last term describes spontaneous emission. In more detail, the excitation rate
\begin{align}\label{eq:R_lup}
	R_{l,u,p} = \Gamma_u \frac{\Omega_{l,u,p}^2 / \Gamma_u^2}{1+4 \delta_{l,u,p}^2 / \Gamma_u^2}
\end{align}
describes the transition strength between the upper and lower level, $u$ and $l$, for a given laser contribution $p$. The Rabi frequency
\begin{align}\label{eq:Rabifreq_lup}
	\Omega_{l,u,p} = \Gamma_u \sqrt{\frac{s_{l,u,p}}{2}} \mel*{u}{\Op{d} \cdot \vec{\epsilon}_p }{l},
\end{align}
characterizes the interaction strength for the specific transition driven on resonance. The parameters appearing in this expression are given relative to those of an equivalent two-level system with the same transition frequency, where the intensity and the Rabi frequency are connected by the simple relation $I/I_\text{s} = 2 \Omega^2/\Gamma^2$. Here, $I_\text{s}$ denotes the saturation intensity of the transition. The strength of an individual transition is taken into account in the matrix elements using the electric dipole moment operator $\Op{d}$.
The saturation parameter $s_{l,u,p} = I_p/I_{\text{s};l,u}$
is related to the intensity $I_p$ of the laser component $p$ and the respective saturation intensity $I_{\text{s};l,u}=\pi hc\Gamma_u/3\lambda_{l,u}^3$ of a transition. Here, $\lambda_{l,u}=2\pi c/\omega_{l,u}$ denotes the wavelength of the transition. In addition, for every light field $p$, we define a detuning $
	\delta_{l,u,p} = \omega_p - \omega_{l,u} - \vec{k}_p \cdot \vec{v}, $
where the term $-\vec{k}_p \cdot \vec{v}$ is the Doppler shift of a particle moving with velocity $\vec{v}$. 

The branching ratios $r_{l,u}$, describe the probability of spontaneous decay from an excited state $u$ to a particular ground state level $l$. These branching ratios are normalized, i.e., $\sum_l r_{l,u} = 1$, to ensure a constant total population $\sum_l \Nl_l + \sum_u \Nu_u$ in the time evolution of the rate equations.

Several important quantities can be derived from the solutions of the rate equations. The photon scattering rate can be found from
\begin{align}\label{eq:Rsc=GammaNe}
	R_\text{sc} = \sum_u \Gamma_u \Nu_u.
\end{align}
From this, the number of scattered photons in the time interval $[t_1, t_2]$ is evaluated as $n_\gamma = \int_{t_1}^{t_2} R_\text{sc} dt$.

The mean force acting on the particle is obtained by taking into account the momentum change due to the absorption and stimulated emission events,
\begin{align}\label{eq:F_rateeqs}
	\vec{F} = \hbar \sum_{l,u,p} \vec{k}_p R_{l,u,p} \qty(\Nl_l - \Nu_u) .
\end{align}

Note that the random walk due to spontaneous emission events averages out over many optical cycles and is not included in this rate equation model.

In addition to the interaction with the light fields, the effects of an applied external static magnetic field can be qualitatively taken into account in the rate equations.
The most important aspect for simulating optical cycling is Larmor precession, which mixes dark with bright ground states. This is qualitatively modeled by adding the term
\begin{align} \label{eq:magnremixMatrix}
	-\sum_{l'} M_{l,l'} \qty( \Nl_l	-\Nl_{l'})
\end{align}
to the rate equations in Eq.~\ref{eq:rateeqs_l}. In this expression a magnetic field is imitated by an empirical mixing matrix $M_{l,l'}$, which transfers populations between neighboring sublevels of individual hyperfine states~\footnote{The minus sign in this expression is chosen by convention, so that population is removed from a driven level and transferred into dark magnetic sublevels by the remixing.}. Thus, the elements of $M_{l,l'}$ are only nonzero if the two ground states $l$ and $l'$ are neighboring magnetic sublevels with $m_F'-m_F = \{0,\pm 1\}$ belonging to the same hyperfine level $F$. The remixing strength is adjusted using the magnitude of nonzero values, which are typically on the order of the scattering rate $\Gamma$, to observe a pumping effect among the sublevels.

The rate equations can provide good estimates of population dynamics, branching losses, scattering rates, and Doppler forces.
However, they neglect all coherent effects due to absorption and stimulated emission processes. In particular, it is well known that, in the regime of low velocities and small magnetic field strengths, Sisyphus-type forces can occur, which are not captured in the rate equation model. To properly account for all these effects, as well as internal state dynamics induced by external magnetic fields, the optical Bloch equations are introduced in the following.

\subsection{Optical Bloch equations}
The optical Bloch equations present a complete picture of an atom or molecule interacting with a light field. Following the work of Gordon and Ashkin~\cite{Gordon1980}, who derived the optical force equations for calculating particle motion, Ungar \etal~\cite{Ungar1989} expanded this theory to include multi-level atoms. Recently, Devlin and Tarbutt applied these equations to analyze three-dimensional laser fields~\cite{Devlin2016} and to model molecular cooling~\cite{Devlin2018}.

For our modeling, we follow the approach of Ref.~\cite{Devlin2018} and include both the effects of the electric light field $\Op{E}$ and an external magnetic field $\Op{B}$ through the interaction with the respective electric and magnetic dipole moments $\Op{d}$ and $\Op{\mu}$. Atomic or molecular eigenstates are denoted as $\ket{e/g,F_a,m_a}_\text{s}=\ket{e/g,a}_\text{s}$, where the subscript $s$ denotes a time-independent eigenstate in the Schr\"odinger picture. These eigenstates belong to an electronically excited or ground state $e$ or $g$, and are characterized by angular momenta $F_a$, their projection along the $z$ axis $m_a$, and the energy $\hbar \omega_{e/g,a}$ labeled by an index $a$. Operators and their time dependence are expressed in the Heisenberg picture 
\begin{align}
	\ket{e/g,a}\bra{e/g,b}\notag =
	e^{-i\omega_{e/g,a}t} \ket{e/g,a}_\text{s} \,
	 e^{+i\omega_{e/g,b}t} \bra{e/g,b}_\text{s}.
\end{align}
With this, the optical Bloch equations in the rotating-wave approximation are given by~\cite{Devlin2018}
\onecolumngrid
\begin{align}\label{eq:OBEsa}
	\dv{ \densM{g,a}{e,b} }{\tau} 
	= &\sum_c \densM{g,a}{g,c} \sum_p \tilde{\Omega}_{c,b,p}\,
		e^{-i\tau \tilde{\delta}_{c,b,p}} \notag\\
	&-\sum_{c'} \densM{e,c'}{e,b} \sum_p \tilde{\Omega}_{a,c',p}\,
		e^{-i\tau \tilde{\delta}_{a,c',p}} \notag\\
	&+ i \sum_n \densM{g,F_a,m_a}{e,F_b,n}
		\sum_q (-1)^q \beta_q \melS{e,F_b,m_b}{\bar{\mu}_{-q}}{e,F_b,n} \notag\\
	&- i \sum_m \densM{g,F_a,m}{e,F_b,m_b}
		\sum_q (-1)^q \beta_q \melS{g,F_a,m}{\bar{\mu}_{-q}}{g,F_a,m_a} \notag\\
	&- \frac{1}{2} \tilde{\Gamma}_b \densM{g,a}{e,b} ,
\end{align}

\begin{align}\label{eq:OBEsb}
	\dv{ \densM{e,a}{e,b} }{\tau} 
	= &\sum_c \densM{e,a}{g,c} \sum_p \tilde{\Omega}_{c,b,p}\,
		e^{-i\tau \tilde{\delta}_{c,b,p}} \notag\\
	&+\sum_{c} \densM{g,c}{e,b} \sum_p \tilde{\Omega}_{c,a,p}^*\,
		e^{+i\tau \tilde{\delta}_{c,a,p}} \notag\\
	&+ i \sum_n \densM{e,F_a,m_a}{e,F_b,n}
		\sum_q (-1)^q \beta_q \melS{e,F_b,m_b}{\bar{\mu}_{-q}}{e,F_b,n} \notag\\
	&- i \sum_m \densM{e,F_a,m}{e,F_b,m_b}
		\sum_q (-1)^q \beta_q \melS{e,F_a,m}{\bar{\mu}_{-q}}{e,F_a,m_a} \notag\\
	&- \frac{1}{2} (\tilde{\Gamma}_a+\tilde{\Gamma}_b) \densM{e,a}{e,b} ,
\end{align}
\begin{align}\label{eq:OBEsc}
	\dv{ \densM{g,a}{g,b} }{\tau} 
	= &-\sum_{c'} \densM{e,c'}{g,b} \sum_p \tilde{\Omega}_{a,c',p}\,
		e^{-i\tau \tilde{\delta}_{a,c',p}} \notag\\
	&-\sum_{c'} \densM{g,a}{e,c'} \sum_p \tilde{\Omega}_{b,c',p}^*\,
		e^{+i\tau \tilde{\delta}_{b,c',p}} \notag\\
	&+ i \sum_n \densM{g,F_a,m_a}{g,F_b,n}
		\sum_q (-1)^q \beta_q \melS{g,F_b,m_b}{\bar{\mu}_{-q}}{g,F_b,n} \notag\\
	&- i \sum_m \densM{g,F_a,m}{g,F_b,m_b}
		\sum_q (-1)^q \beta_q \melS{g,F_a,m}{\bar{\mu}_{-q}}{g,F_a,m_a} \notag\\
	&+ \sum_{c',c''} e^{-i\tau ( \tilde{\delta}_{a,c',0} - \tilde{\delta}_{b,c'',0}) }
		\frac{1}{2} (\tilde{\Gamma}_{c'}+\tilde{\Gamma}_{c''})  \densM{e,c'}{e,c''} \notag\\
		&\qquad \times\sum_q \melS{e,c'}{\hat{d}_q}{g,a} \melS{g,b}{\hat{d}_q}{e,c''} .
\end{align}
\twocolumngrid

In each of these three equations, the first two terms correspond to the coherent atom-light interaction, the next two include the effect of an external magnetic field, and the last ones correspond to spontaneous emission.
The indices $c'$ and $c''$ represent all excited states and $c$ all ground states, the projection number $q \in \{0,\pm 1\}$ labels the tensor components in the spherical basis, and $p$ again denotes the laser components. The sums in the magnetic field terms run over different sets of values $-F_a \le m \le F_a$ and $-F_b \le n \le F_b$, which depend on the angular momentum numbers $F$ of the states involved. Spontaneous emission is treated by the \mbox{radiation reaction approximation}~\cite{Ungar1989}.

As dimensionless times $\tau = \bar{\gamma} t$ with an arbitrary linewidth $\bar{\gamma}$~\footnote{The arbitrary linewidth $\bar{\gamma}$ is typically chosen as the linewidth $\Gamma$ of one of the electronic states.} are used, the natural linewidth, detuning, and Rabi frequency are now given by
\begin{align}
    \tilde{\Gamma}_u &= \Gamma_u/{\bar{\gamma}},\\
	\tilde{\delta}_{l,u,p} &= \delta_{l,u,p}/{\bar{\gamma}}, \label{eq:delta_tilde}\\
	\tilde{\Omega}_{l,u,p} &= \frac{i}{2} e^{i\varphi_p} h_{l,u,p} 	
		\frac{\Omega_{l,u,p}}{\bar{\gamma}} . 	\label{eq:Omega_tilde}
\end{align}
The coefficients $h_{l,u,p}$ can take the values $0$ or $1$ depending on whether the driving of the transition \transition{l}{u} due to the laser component $p$ should be included or neglected in the simulation. 

The electric field is treated classically
\begin{align}
	\vec{E}_p(\vec{x},t) = E_{0,p}\, \vec{\epsilon}_p\, e^{i(\vec{k}_p\cdot\vec{v}t + \varphi_p)} \cos(\omega_p t) ,
\end{align}
where $\vec{\epsilon}_p$ is the normalized polarization vector of the laser component $p$ with wavevector $\vec{k}_p$, angular frequency $\omega_p$, and intensity $I_p=c\epsilon_0 E_{0,p}^2 /2$. The expression in the exponent is deduced from the general phase $\vec{k}\cdot \vec{x} + \varphi'$, where the position dependence is assumed to be linear in velocity, $\vec{x}=\vec{v}t+\vec{x}_0$, which in turn is absorbed into the optical Bloch equations via the Doppler shift in the detuning (Eq.~\ref{eq:delta_tilde}). Thus, the corresponding laser phase also incorporates the initial position of a particle through $\varphi_p = \vec{k}_p \cdot \vec{x}_0 + \chi_p$ with an offset $\chi_p$.

The applied magnetic field is assumed to be small enough to produce only linear Zeeman shifts and takes the form of a classical field vector
\begin{align}
	\vec{B} = \frac{\hbar \bar{\gamma}}{\mu_B} \sum_q \beta_q \epsilon_q^* ,
\end{align}
where $\epsilon_q$ are the usual spherical basis vectors and $\mu_B$ is the Bohr magneton. The field amplitude $\beta_q$ is defined in analogy to the electric field amplitude $E_{0,p}$ above. 
Using the effective magnetic $g$-factors $g_F$, the respective matrix elements of the magnetic moment operator $\Op{\mu}$ are
\begin{multline}
	\melS{e/g,F_a,m}{\bar{\mu}_{q}}{e/g,F_a,n} \notag=\\
	-g_{F_a} (-1)^{F_a-m} \sqrt{F_a(F_a+1)(2F_a+1)} \Wjt{F_a}{1}{F_a}{-m}{q}{n} .
\end{multline}
The resulting forces acting on the particles are captured using a force operator, which is derived using the Heisenberg equation of motion
\begin{align}
	\Op{f} &= \dv{\Op{p}}{t} = -\frac{i}{\hbar} \left[ \Op{p},\Ham \right] = -\grad{\Ham}.
\end{align}
The expectation value $\langle\Op{f}\rangle=\vec{F}$ is evaluated for a molecule moving with constant velocity $\vec{v}$ through the light field using the expectation values $\densM{e,c'}{g,c}$ obtained from the optical Bloch equations. The corresponding force is
\begin{align}\label{eq:F_OBEs}
	\vec{F} =2 \hbar \bar{\gamma} \Re\! \qty(
	\sum_{c,c'} \densM{e,c'}{g,c} \sum_p \tilde{\Omega}_{c,c',p}\,
	e^{-i\tau \tilde{\delta}_{c,c',p}} \vec{k}_p ) .
\end{align}
The direct magnetic contribution to the force is  small and can therefore be neglected.

\subsection{Trajectory calculations}
As discussed in the preceding sections, the rate model and the optical Bloch equations constitute powerful tools to obtain the equilibrium force and  internal state dynamics of atoms and molecules. However, in realistic simulations of a laser cooling experiment, the trajectory of a particle as it passes through regions of different laser beam configurations, laser field intensities, and external fields must be taken into account to derive its time-dependent velocity.

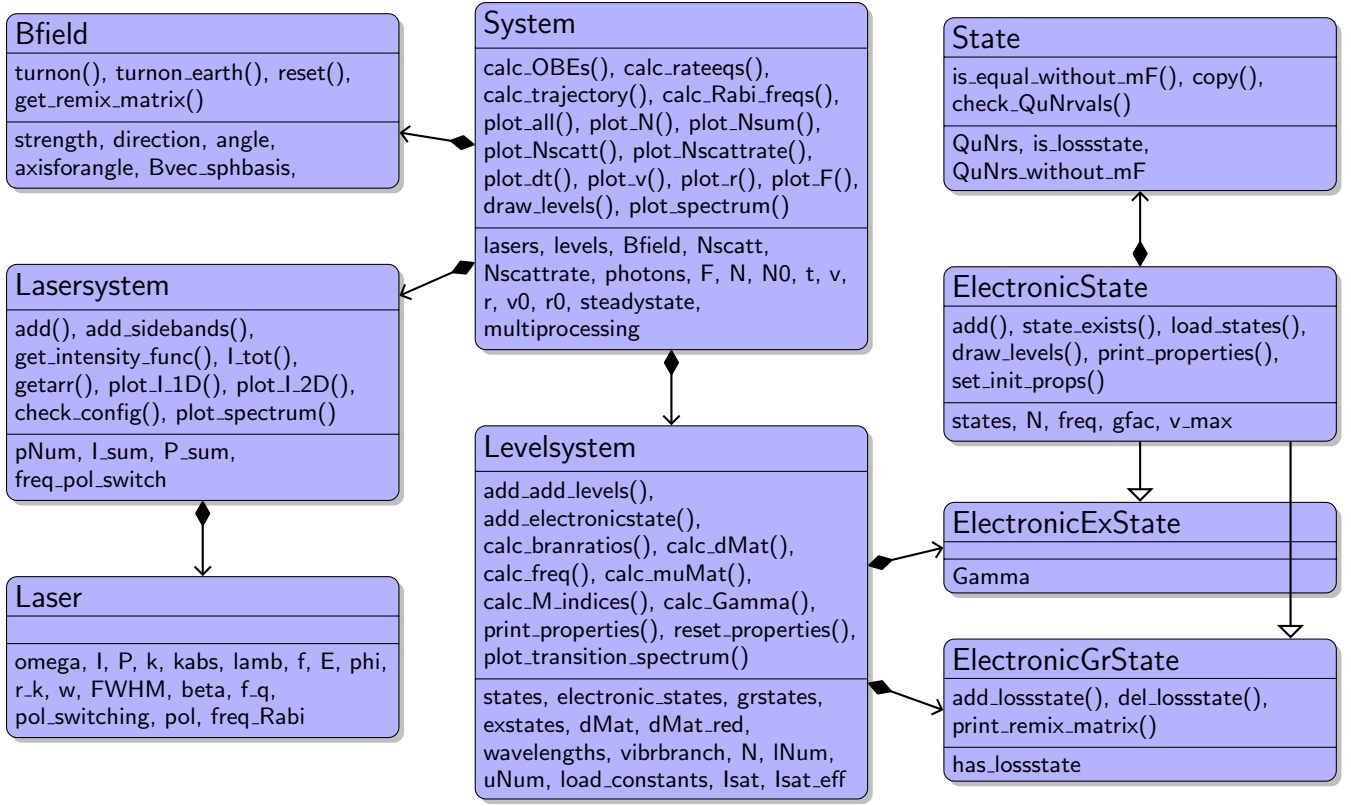
\begin{figure*}[tb]
	\small
	\input{CodeStructure}
	\caption{Core classes, their relationships, important methods, properties, and attributes of the Python package \Py{MoleCool}. This object-oriented simulation toolbox enables simulating internal dynamics of multi-level systems interacting with a laser light and magnetic field via the rate equations and the OBEs. Class composition and inheritance is denoted by arrows with diamonds and open tips, respectively. Methods are indicated by parenthesis.
	}
    	\label{fig:CodeStructure}
\end{figure*}

In general, this leads to a position- and velocity-dependent force $\vec{F}(\vec{x},\vec{v})$ acting on the particle. The effect of this force can be propagated through time and space with the equations of motion
\begin{subequations}\label{eq:EqsofMotion}
\begin{align}
	\dv{\vec{r}}{t} & = \vec{v} , \\
	\dv{\vec{v}}{t} &= \frac{\vec{F}(\vec{x},\vec{v})}{m} .
\end{align}
\end{subequations}
In the case of the rate equations,  $\vec{F}(\vec{x},\vec{v})$ can simply be substituted by the expression of the force given in Eq.~\ref{eq:F_rateeqs}, without introducing substantial computational overhead. This results in a velocity-dependent scattering rate $R_{l,u,p}(\vec{v})$ (see Eq.~\ref{eq:R_lup}) due to the Doppler shift. The position-dependence in $R_{l,u,p}(\vec{x},\vec{v})$ due to a certain laser beam configuration is included by assigning each laser component $p$ a Gaussian intensity distribution $I_p(r)$ dependent on the radial distance $r$, power $P_p$ and $1/e^2$ radius $w_p$,
\begin{align} \label{eq:I(r)-distr}
	I_p(r) = \frac{2P_p}{\pi w_p^2} \exp(-\frac{2r^2}{w_p^2}) .
\end{align}
More generally, the total intensity is given by $I_\text{tot}(\vec{x}) = \sum_p I_p(\vec{x})$ for laser components placed at arbitrary positions with specific wavevectors. In addition, the beams can be widened in one transversal direction by the $1/e^2$ radius $\tilde{w}_p$ that yields the intensity $I_{p,\text{max}}=\frac{2P_p}{\pi w_p\tilde{w}_p}$.

Using the rate equations, it is possible to calculate a particle's propagation $(\vec{x}(t), \vec{v}(t))$ through a certain laser beam configuration $I_\text{tot}(\vec{x})$ with comparably low computational effort in a few seconds on a personal computer. This makes Monte Carlo simulations possible, where the behavior of particle ensembles is modeled by sampling many trajectories with different initial conditions $(\vec{x}(t_0),\vec{v}(t_0))$. These initial conditions can, for example, be chosen from a Boltzmann distribution describing a particle ensemble with a particular temperature. 

In contrast to the rate equations, in the case of the optical Bloch equations, the computational time is significantly higher and scales quadratically with the number of levels $N$ involved. This prohibits direct Monte Carlo simulations and single-particle propagation through a long interaction region with varying intensity. Thus, another approach has to be used in order to calculate trajectories. Specifically, we evaluate the optical Bloch equations for a particle moving with constant velocity in a uniform light field until the periodic quasi-steady state is established. This is repeated for many different intensities, velocity directions, absolute velocities, laser phases, laser detunings, and magnetic field vectors, which sets up a large parameter space out of which the force $\vec{F}(\vec{x},\vec{v})$ is constructed subsequently. Using this force, the equations of motion given in Eq.~\ref{eq:EqsofMotion} can be used to classically calculate a large number of single-particle trajectories.

\section{Setting up a simulation}
The \Py{MoleCool} toolbox is an object-oriented Python simulation software to model the dynamics described in the previous chapters. The diagram in Fig.~\ref{fig:CodeStructure} shows all core classes of this code with their most important methods, properties,  attributes, and relations to each other. 

A key feature of the code is its ability to easily construct an arbitrary multi-level system, together with a complex light field comprising several frequency components and laser beams, as well as an external magnetic field. For a given setup, the light field can drive multiple transitions characterized by their relative energies and branching ratios. Based on this description, the toolbox evaluates the internal particle dynamics using either rate equations or optical Bloch equations.

The resulting initial value problem is solved using the \Py{solve_ivp} module of \Py{scipy}. Because the differential equation function is evaluated many times during integration, the computational cost can be reduced significantly by translating this function into optimized machine code. This is accomplished with the \Py{numba} package, whose open-source just-in-time (JIT) compiler enables Python functions to run at speeds approaching those of C or FORTRAN code.

In the case of rate equations, even the calculation of single-particle trajectories over long distances through multiple laser beams with Gaussian intensity profiles remains computationally inexpensive. Using the \Py{LSODA} integration method with variable step size provides the desired performance. This efficiency, in particular, facilitates Monte Carlo simulations that track the motion of particles through realistic intensity patterns formed by multiple laser beams.

In contrast, solving the optical Bloch equations benefits primarily from the standard explicit Runge–Kutta method \Py{RK45} of order 5. Modeling comparable trajectories requires first calculating the quasi-steady-state force over a wide range of parameters, which is considerably more time-consuming. This process is accelerated by parallelizing the simulations across multiple cores, with each parameter configuration evaluated on a separate core. For this purpose, we employ Python’s \Py{multiprocessing} module.

\section{Basic examples}
In the following, we demonstrate the capabilities of our toolbox with a series of examples of increasing complexity. Beyond their practical purpose, each example highlights specific aspects of light–matter interactions that are essential for understanding laser cooling forces in realistic molecules. All examples are provided in the package's online repository~\cite{GithubLangenGroup}, together with step-by-step implementation guides.

\begin{figure}[tb]
	\centering
	\includegraphics[width=0.95\linewidth]{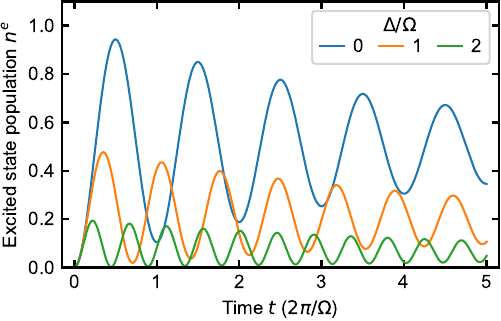}
	\caption{Rabi oscilations in a simple two-level system for three different detunings $\Delta$. The damping of the oscillation amplitude is a result of the finite lifetime of the excited state.
    On resonance ($\Delta = 0$), the excited-state population approaches its maximum steady-state value of $\Nu = 1/2$. For finite detuning, the oscillation frequency increases while the average scattering rate $\Gamma \Nu$ decreases.}
	\label{fig:twolevel}
\end{figure}

\subsection{Two-level Rabi oscillations}
We start by modeling the coherent dynamics in a textbook-like two-level system, i.e., a system with $N_\text{g}=1$ and $N_\text{e}=1$, often also referred to as 1+1 levels, using the optical Bloch equations. If such a system is driven by an external laser, it undergoes coherent Rabi oscillations, where the excited state population oscillates with the generalized Rabi frequency $\sqrt{\Omega^2+\Delta^2}$. Here, $\Omega$ is the resonant Rabi frequency and $\Delta$ is the detuning of the laser field from the transition frequency of the two-level system.

When the finite lifetime of the excited state is taken into account, the oscillations are damped and converge to a steady state with the scattering rate $R_\text{sc}=\Gamma \Nu$. This rate saturates at the maximum possible value of $\Gamma/2$, or at lower values for finite detuning. Examples of the resulting time evolutions, calculated from the OBEs for different parameter sets, are shown in Fig.~\ref{fig:twolevel}.

\begin{figure}[tb]
	\centering
	\includegraphics[width=0.95\linewidth]{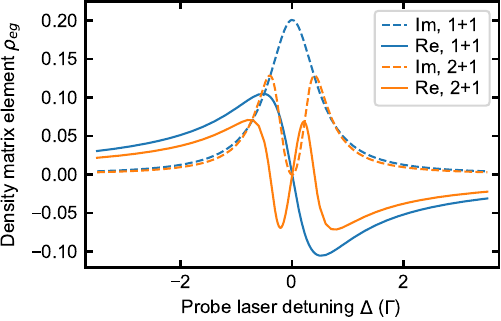}
	\caption{Real (dispersion) and imaginary part (absorption) of the density matrix element (susceptibility) in comparison for a simple 1+1 two-level system and a 2+1 three-level system demonstrating the effect of electromagnetically-induced transparency. The pump laser in the 2+1 system is assumed to be on resonance with its pump transition and only the probe laser is detuned.}
	\label{fig:threelevel}
\end{figure}

\subsection{Electromagnetically-induced transparency\\in a three-level system}
In the next example, we increase the number of ground state levels to $N_\text{g}=2$. This leads to a $2+1$ or $\lambda$- system~\cite{CohenTannoudji,Fleischhauer2005}, where a single excited state ($N_\text{e}=1$) couples to both ground states. The resulting two transitions are commonly referred to as \emph{probe} and \emph{pump}, respectively. 

The absorption and dispersion of an atomic or molecular gas characterized by such three-level systems are conveniently represented by the imaginary and real parts of the susceptibility~\cite{CohenTannoudji}. This susceptibility is proportional to the off-diagonal matrix element $\rho_{eg}$ of the density matrix, which can be derived from the solution of the OBEs. 

In our example, we keep the pump laser on resonance and derive the susceptibility as a function of the probe laser detuning $\Delta$. 

Fig.~\ref{fig:threelevel} compares our numerical results for the 2+1 system with those of a two-level system. While absorption in a two-level system is maximized for a resonant laser beam, a three-level system behaves differently. Here, constructive interference between the two possible ground-to-excited-state transition pathways gives rise to dark states~\cite{Fleischhauer2005}. Consequently, a transparency window emerges at vanishing detuning of the probe laser, manifesting the phenomenon of electromagnetically induced transparency (EIT)~\cite{Fleischhauer2005}.

\begin{figure}
	\centering
    \includegraphics[width=\linewidth]{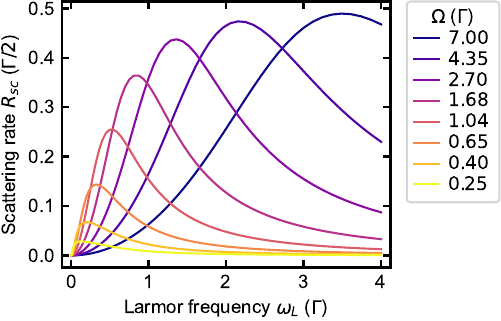}
	\caption{Steady-state scattering rate $R_\text{sc}$ as a function of Larmor precession frequency $\omega_L$ for several Rabi frequencies $\Omega$ in a $3+1$ system with excited state lifetime $\Gamma$. In such a type-II transition, the usual saturation effect that manifests in $R_\text{sc}$ for high intensities (or, equivalently, high $\Omega$) disappears. For each Rabi frequency,  $R_\text{sc}$ peaks at a distinct magnetic field strength (or, equivalently, $\omega_L$), but then strongly decays. This behavior is typical for a system with dark state remixing.}
	\label{fig:Fig4_3+1system_B-field}
\end{figure}

\subsection{Optical cycling on a type-II transition}
Dark states play an important role in multi-level systems, whether they are induced by coherent coupling of states or from the absence of a laser driving certain transitions. To illustrate this importance, in the next example, we consider a configuration with $N_\text{g}=3$ ground states coupled to $N_\text{e}=1$ excited state. Such systems, where the ground-state manifold contains more sublevels than the excited-state manifold, are referred to as type-II transitions and naturally give rise to dark states.

Typical examples of type-II transitions in atoms and molecules are those between hyperfine levels with $F=1$ in the ground state and $F'=0$ in the excited state. When such a system is driven with light of a given---e.g., linear---polarization, only one of the three magnetic sublevels in the ground state couples to the single sublevel of the excited state. In contrast, spontaneous decay from the excited state can populate all three ground-state sublevels. As a result, population gradually accumulates in the uncoupled dark states, and scattering eventually ceases. A straightforward way to return these populations to the optical cycle is to couple the ground-state magnetic sublevels with a magnetic field tilted relative to the linear polarization of the driving laser~\cite{Tarbutt2018}. 

Examples of resulting steady-state configurations, obtained from OBE calculations, are shown in Fig.~\ref{fig:Fig4_3+1system_B-field} for several driving Rabi frequencies $\Omega$. In this example, the ground-state $g$-factor is set to unity, and the angle between the magnetic field and the laser polarization is fixed at \SI{45}{\degree}.

We observe that---counter-intuitively---for such a multi-level system a higher intensity does not necessarily lead to more scattering. Instead, for a given intensity, the magnetic field strength has to be carefully chosen to reach the maximum of the photon scattering rate. At this maximum, the Rabi frequency is comparable to the Larmor frequency $\omega_L$, leading to well-matched timescales for excitation and dark-state remixing. This effect is characteristic of type-II transitions.

\label{sec:multileveldynamics}
These observations can be generalized to generic multi-level systems. Under the assumption that all detunings $\delta_{l,u,p}$ are equal, all laser components have equal intensity with a total intensity $I_\text{tot} = \sum_p I_p$, and all excited states are equally populated, the equilibrium scattering rate can be approximated as~\cite{Fitch2021}
\begin{align}\label{eq:R_sceff}
	R_\text{sc} = \frac{\Gamma_\text{eff}}{2} \frac{s_\text{eff}}{1+ s_\text{eff} + 4\delta^2 / \Gamma^2}.
\end{align}
This expression has the same form as the scattering rate in a two-level system, but with the effective parameters
\begin{align}
	\Gamma_\text{eff} &= \frac{2N_\text{e}}{N_\text{g}+N_\text{e}}\Gamma ,\label{eq:Gammaeff}
	\\
	s_\text{eff} &= \frac{I_\text{tot}}{I_{\text{s,eff}}} ,
	\\
	I_{\text{s,eff}} &= \frac{2N_\text{g}^2}{N_\text{g}+N_\text{e}} I_\text{s} . \label{eq:I_sateff}
\end{align}
As a consequence, for high intensity $s_\text{eff} \gg 1$, the maximum scattering rate in Eq.~\ref{eq:R_sceff} is given by $R_\text{sc,max} = \Gamma_\text{eff}/2$, where the populations of all levels $l,u$ are equally populated. This yields a total excited-state population $\Nu = \sum_u \Nu_u = N_\text{e} / (N_\text{g}+N_\text{e})$. 

Level schemes used in molecular laser cooling usually contain more ground states than excited states, a consequence of the parity selection rules employed to suppress rotational branching. As evident from the above equations, the maximum scattering rate is therefore typically well below the familiar two-level limit of $\Gamma/2$. Moreover, additional ground states from higher vibrational or rotational levels must often be included for an accurate description of the dynamics. This further reduces $R_\text{sc,max}$ while simultaneously increasing the effective saturation intensity $I_\text{s,eff}$. For example, in the case $N_\text{e}=1$ and $N_\text{g}=3$ (the 3+1 system), one obtains $R_\text{sc} = 0.5\,\Gamma/2$, in agreement with the asymptoptic value observed in our numerical solution (see Fig.~\ref{fig:Fig4_3+1system_B-field}) and equal to half the two-level value.

\section{Species-specific examples}

We now turn to concrete examples involving molecular species of current experimental interest. The toolbox includes the relevant branching ratios and level schemes for these species. We begin with simple photon-scattering, optical-cycling, and detection experiments, then consider cooling and slowing forces, and finally model coherent bichromatic forces. The example molecules, which range from barium monofluoride (BaF) and ytterbium monohydroxide (YbOH) to radium and calcium monofluoride (RaF and CaF, respectively), exhibit broadly similar but subtly distinct level structures. These examples illustrate how even small differences in level structure must be carefully accounted for when optimizing scattering rates, optical forces, or cooling schemes.

\subsection{Optical cycling of a 12+4 level system in \BaFb}

\begin{figure}[tb]
	\centering
	\includegraphics[width=\linewidth]{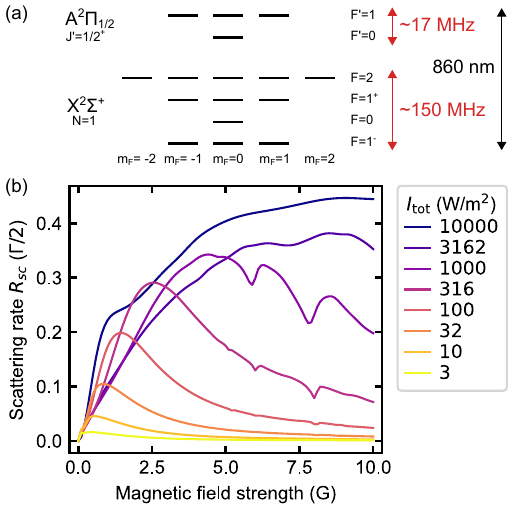}
	\caption{(a) Example $12+4$ level system. Specific parameters are shown for BaF molecules, 
but the structure of the rotationally-closed \gs($N=1$) $\rightarrow$ \exs($J=1/2^+$) transition is the same 
for all other common monohalides considered for laser cooling. 
(b) Steady-state scattering rate $R_\text{sc}$ as a function of magnetic-field strength $B$ 
for several cooling-laser intensities $I_\text{tot}$ in this level system. As in Fig.~\ref{fig:Fig4_3+1system_B-field}, the usual saturation of $R_\text{sc}$ at high intensities disappears for certain magnetic field strengths. In addition, sharp dips in the scattering rate emerge. These dips are reminiscent of EIT, occurring when the Larmor frequency matches the laser detuning from a power-broadened neighboring transition.}
	\label{fig:BaF_Ne_Bfield-I-3angles}
\end{figure}

To make the discussion of multi-level systems more concrete, we first consider a $12+4$ level system. This configuration is typical of rotationally closed \gs($N=1$) $\rightarrow$ \exs($J=1/2^+$) transitions in alkaline-earth monohalides such as CaF, SrF and BaF~\cite{Fitch2021} (see Fig.~\ref{fig:BaF_Ne_Bfield-I-3angles}a). Here, $N$ and $J$ denote the rotational angular momentum and the total electronic angular momentum 
(excluding nuclear spin), respectively. 
The given values yield a rotationally-closed optical cycle~\cite{Fitch2021}. 

In the following, we use branching ratios and energy splittings for \BaFb. For this and all subsequent examples, level structures and branching ratios can be conveniently visualized using the function \Py{draw_levels()} provided in our package.

We solve the dynamics in these systems using the OBEs and find the steady-state scattering rate $R_\text{sc}$ as a function of the magnetic field. In this process, we include several laser frequency components, commonly referred to as \emph{sidebands}, to address all hyperfine states. The results are depicted in Fig.~\ref{fig:BaF_Ne_Bfield-I-3angles}b for different intensities $I_{\text{tot}}$. 

In a $12+4$ level system, the theoretically possible scattering rate remains 
$R_\text{sc} = 0.5\,\Gamma/2$, despite the larger total number of states. 
This follows from the identical ratio of excited to ground states as in the $3+1$ case. 
However, in  full simulations, this limit is reached only asymptotically and only at very high intensities, 
consistent with the significantly increased theoretical saturation intensity of 
$I_\text{s,eff} = 18\,I_\text{s}$.

Overall, the observed behavior is reminiscent of the magnetic-field dependence shown 
for a $3+1$ level system in Fig.~\ref{fig:Fig4_3+1system_B-field}. 
We again find that, for a given intensity $I_{\text{tot}}$, the magnetic field strength 
must be carefully tuned to maximize the photon-scattering rate. 
At large magnetic fields, the scattering rate decreases due to increasingly rapid 
Larmor precession, which is no longer synchronized with the optical cycling rate, 
as well as due to large Zeeman shifts that detune the $m_F$ sublevels from the laser sidebands 
used to address them. 

In addition, sharp dips in the scattering rate—reminiscent of EIT—appear at specific magnetic-field strengths. 
These resonances, previously studied in a $(2\times3)+1$ model system~\cite{BarryPhD}, 
occur when the Larmor frequency matches the laser detuning of a power-broadened neighboring transition.

\subsection{Optical cycling in ${}^{171}$YbOH molecules}
\label{sec:YbOH}

There has recently been important progress in the direct laser cooling and manipulation of polyatomic molecules~\cite{Augenbraun2023}. As the next example we therefore model optical cycling in ytterbium monohydroxide (YbOH). In particular, the odd isotopologues  \mol{173}{YbOH} and \mol{171}{YbOH} are highly sensitive probes for precision measurements of nuclear properties and symmetries~\cite{Kozyryev2017}. 

\begin{figure}[tb]
    \includegraphics[width=\columnwidth]{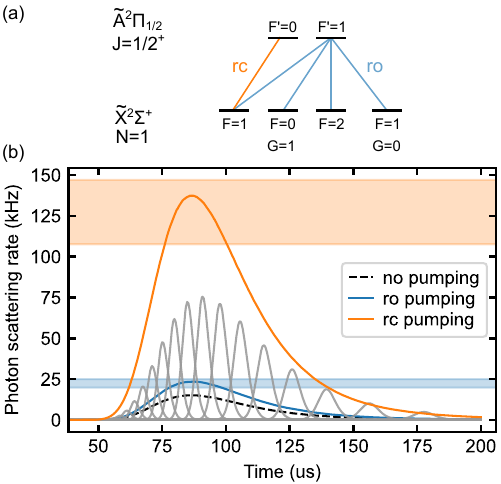}
	\caption{(a) Strongly simplified level scheme of the  \transition{\tilde{A}^2\Pi_{1/2}}{\tilde{X}^2\Sigma^+} in \mol{171}{YbOH}, neglecting additional vibrational states, further substructure arising from the non-zero hydrogen nuclear spin, and the magnetic sublevels taken into account in the simulation. The full level structure includes $24+8$ states per vibrational mode. The $\tilde{A}(F'=0)$ and $\tilde{A}(F'=1)$ states decay into a single and all ground state levels, respectively. This results in rotationally closed (rc) and rotationally open (ro) transitions, respectively~\cite{Zeng2023}. 
    (b)~Monte Carlo-based modeling of optical cycling in \mol{171}{YbOH} using rate equations. Individual molecules are first propagated through a pump laser beam driving either rc or ro transitions. They are subsequently probed with a weaker beam resonant to the rc transition on the first repumper, with the corresponding photon scattering rates representing the resulting fluorescence (example traces in gray). The time-of-flight signal of the entire molecular cloud (colored lines), obtained by summing over all trajectories and compared to the case without a pumping beam, shows excellent agreement with the experimental data of Ref.~\cite{Zeng2023}. 
The horizontal shaded regions indicate the experimental values, with their widths corresponding to the measurement uncertainties.
}
	\label{fig:yboh}
\end{figure}

Motivated by this fact, the optical cycling of these isotopologues has recently been demonstrated for the first time~\cite{Zeng2023}. In the following, we model the observed optical cycling signals. For our demonstration, we use the \transition{\tilde{X}^2\Sigma^+(N=1)}{\tilde{A}^2\Pi_{1/2}(J'=1/2^+)} transition of \mol{171}{YbOH}, with the relevant substructure of this transition summarized in Fig.~\ref{fig:yboh}a. Cycling in this species is complicated by the nonzero nuclear spins of both the \mol{171}{Yb} and \mol{}{H} nuclei, leading to a complex hyperfine structure with many states. For our simulations, the level structure is derived from a recently measured set of molecular spectroscopic constants~\cite{Pilgram2021,Zhang2021} and includes 24+8 states per vibrational mode. We incorporate the vibrational ground ($\nu=0$) and first excited ($\nu=1$) states, which are essential for the experimentally realized pumping and probing schemes~\cite{Zeng2023}. 
Additionally, we also include a single effective loss channel to account for all residual population loss channels. In addition to higher-order vibrational branching losses, these include contributions from 4$f$-hole states of the Yb atom (which are well known to also occur in YbF molecules, where they have been studied in detail~\cite{Popa2024}), as well as branching into higher rotational states due to excited-state hyperfine-induced mixing~\cite{Pilgram2021,Kogel2025spectroscopy}. This simplified modeling is justified by the low total number of scattered photons expected in the observed optical cycling, but it would no longer be suitable when describing laser cooling in YbOH. Also, for modeling of this optical pumping and the subsequent probing, the rate equations are sufficient, as coherent effects do not play any significant role. 

Two different types of transitions can be addressed by selective excitation of the $\tilde{A}(F'=0)$ and $\tilde{A}(F'=1)$ states in ${\tilde{A}^2\Pi_{1/2}(J'=1/2^+)}$, respectively, which are split by approximately \SI{400}{\mega\hertz} in \mol{171}{YbOH}. Here, $F$ denotes the total angular momentum, excluding the unresolved hydrogen nuclear spin. In the first case, rotationally closed (rc), the $F'=0$ state is selectively driven and decays with a probability of \SI{99}{\percent} back to a single ground state $\tilde{X}(G=1,F=1)$, thus maintaining a nearly closed cycle. In contrast, excitation of the rotationally open (ro) $F'=1$ state results in decay into all accessible hyperfine components of $\tilde{X}(G=1,F=0,1,2)$ and $\tilde{X}(G=0,F=1)$ with comparable probabilities, with $G$ denoting an intermediate angular momentum. As a result, the population is quickly lost to dark states, leading to termination of the optical cycle.

As in the experiment, we compare the population transfer from the vibrational mode $\nu=0$ to $\nu=1$, when the pump drives closed rc or open ro transitions. This comparison is realized by operating the probe on an rc transition of the $\nu=1\rightarrow\nu'=0$ repumper.

To allow a direct comparison with the experimental results~\cite{Zeng2023}, we simulate the trajectories of 100 individual molecules with distinct initial forward velocities. Each molecule first propagates through a $5$-mm-wide pump beam with $10\,$mW of power and subsequently crosses a weaker 
$100\,$\textmu W probe beam. The photon scattering rates of the individual molecules in the probe beam directly represent the fluorescence signal. The ensemble time-of-flight (TOF) signal is obtained by summing the scattered photons of all simulated molecules, weighted using a Boltzmann distribution based on their initial velocities, assuming a mean forward velocity of $180\,$m/s. 

A summary of the simulation is presented in Fig.~\ref{fig:yboh}b. The results show enhancements of the fluorescence by approximately \SI{55}{\percent} and \SI{800}{\percent} for rotationally open and closed pumping, respectively, relative to the case where no pump laser is applied (dashed black line). These values are in excellent agreement with the experimental measurements~\cite{Zeng2023}, as represented by the horizontal bars in the figure.

\subsection{Frequency-chirped slowing of a ${}^{138}$BaF beam}
\label{sec:bafslowing}
After demonstrating the optical cycling of molecules in a beam, we will next model the beam's deceleration using the radiation-pressure force. A slow molecular beam is essential for extending both the interaction time of molecules for precision beam experiments and for facilitating their efficient loading into a magneto-optical trap (MOT) for subsequent laser cooling~\cite{Langen2023,Tarbutt2018}.

Slowing heavy molecules is particularly challenging due to their small recoil velocity and the reduced scattering force associated with type-II transitions, as discussed above. Also, a Zeeman slower, the standard technique typically used to efficiently slow down atomic beams, is challenging to implement for molecules~\cite{Petzold2018,Liang2019,Sawaoka2023}. 
Experiments thus typically use frequency-chirped laser slowing to produce a compressed and velocity-controlled molecular beam. In the following, we simulate this slowing for heavy \BaFb\ molecules. For this, we follow the approach of Truppe \etal~\cite{Truppe2017}, who studied lighter CaF molecules.

At the start of the simulation, the molecules are sampled again over different initial forward velocities. Because the recoil velocity of \BaFb, $v_\mathrm{rec}=2.9\,$mm/s, is relatively small for the cooling wavelength $\lambda_{00}=860\,$nm and linewidth $\Gamma = 2\pi\times 2.84\,$MHz, a large number of photons—on the order of \num{65000}—must be scattered to decelerate the molecules from typical buffer-gas beam velocities of $190\,$m/s~\cite{Albrecht2020} to near rest. Given the Franck–Condon factors of this species~\cite{Rockenhauser2023}, three vibrational repump lasers are required in addition to the cooling laser. Taking rotational and hyperfine substructure into account, but neglecting losses to an intermediate $A'\Delta$ state~\cite{Rockenhauser2023,Rockenhaeuser2024lasercooling}, this results in a $49+12$ level system.

\begin{figure}[tb]
	\centering
	\includegraphics{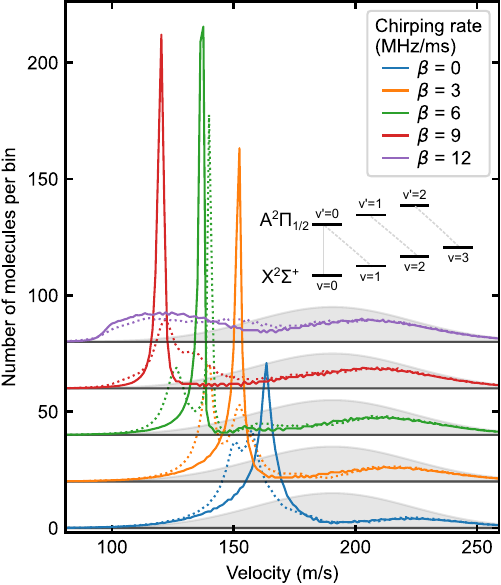}
	\caption{Monte Carlo simulation of radiation pressure frequency-chirped slowing of \BaFb\, molecules for several chirping rates $\beta$ and a slowing time of \SI{9}{\milli\second}. The figure shows the velocity distributions with (colored lines) and without slowing (gray shaded areas) applied. The slowed distributions show a well-defined mean velocity with a highly compressed velocity spread, resulting in an intense, cold, velocity-controlled molecular beam with a high density. Dashed colored lines correspond to equally-spaced hyperfine sidebands created using a single EOM. For the solid lines, a sideband spectrum precisely matching the molecular hyperfine structure was applied, as previously realized for CaF molecules~\cite{Holland2020}. The latter shows significantly better slowing efficiency, highlighting the importance of the choice of sidebands~\cite{Kogel2025optimization}. The inset shows the relevant vibrational level structure of BaF, with one cooling laser (solid line) and three repump lasers (dashed lines), neglecting further substructure, that for each state is identical to the one detailed in Fig.~\ref{fig:BaF_Ne_Bfield-I-3angles}a.
	}
	\label{fig:chirping}
\end{figure}

In the simulation, a large slowing laser beam width is assumed to produce a uniform intensity distribution along the slowing axis with $I_{00}=I_{10}=I_{21}=I_{32}= 7000\,$W/$\textrm{m}^2$, where $I_{\nu \nu'}$ denotes the intensity on the usual \gs($\nu$) $\rightarrow$  \exs($\nu'$) transitions used for cycling. Here, the \emph{cooling laser} addresses the transition with $\nu=0\rightarrow\nu'=0$ and \emph{repump lasers} address
transitions with $\nu\rightarrow\nu'=\nu-1$ (see Fig.~\ref{fig:chirping}). The high intensities on the repumping transitions are chosen to compensate for the unfavorable Franck-Conden overlap of these transitions. 

The slowing light is turned on at $t=0$ for a duration of $9\,$ms. Due to the Doppler shift $\Delta f = v_0/c$, the initial laser frequencies must be detuned to the red to be in resonance with molecules moving with the initial velocity $v_0$ in the direction opposite to the laser beam. Absorption from this beam slows the molecules down and they experience a changing Doppler shift. To account for this, the laser frequencies are chirped linearly with a rate $\beta$ for the cooling laser and $\beta\times\lambda_{00}/\lambda_{\nu\nu'}$ for repump lasers. The resulting amplitudes of each bin of the velocity distribution are derived by weighting molecules with particular initial forward velocities according to a Boltzmann distribution. 

The results of the simulation are shown in Fig.~\ref{fig:chirping}. When no slowing light is used, the velocity distribution has the expected Gaussian shape. With slowing light, the respective final distributions as a function of the chirp rates reveal compression and slowing of the initial molecular beam. For increasing chirping rates $\beta$, the velocity distribution is compressed and shifted to smaller velocities. If the chirp rate is too high $\beta > \SI{9}\,$MHz/ms, the molecules are insufficiently decelerated and cannot follow the chirped laser frequency anymore. 

For illustration, we compare results for a simple choice of equidistant laser sidebands, created with a single electro-optical modulator (EOM), and for optimized sidebands, which match the four strongest hyperfine transitions and lead to significantly more efficient slowing~\cite{Kogel2025optimization}. Similar improvements for optimized laser spectra have also been observed in slowing experiments with CaF~\cite{Holland2020}. 

For the simulated parameters, the width of the initial velocity distribution is reduced by roughly a factor of 10, and the mean velocity decreases by more than $60~\text{m/s}$, from an initial $190~\text{m/s}$ to approximately $125~\text{m/s}$. Achieving complete deceleration to rest would require extending the interaction time by a factor of 3---4 compared to the present simulation. For \BaFb, this corresponds to a slowing distance of approximately \SI{3}{\metre}. In this regime, the transverse divergence of the beam during slowing would also become significant, drastically reducing the fraction of molecules that can be captured for further laser cooling. One strategy to mitigate these limitations is to start from lower initial velocities, achievable, for example, with dual-stage buffer-gas sources and reduced buffer-gas temperatures, as recently demonstrated for YbF~\cite{White2024}. Another approach is to implement larger coherent forces, as discussed in example F.

\begin{figure}
	\centering
	\includegraphics{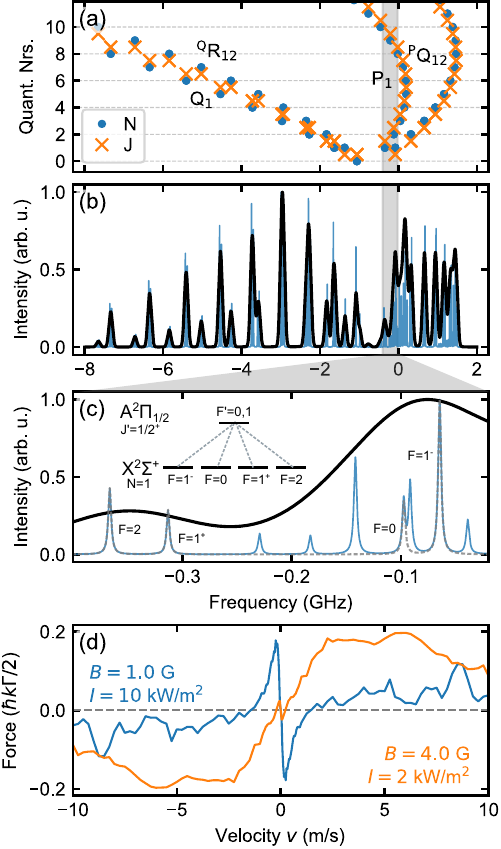}
	\caption{Identification of a laser cooling scheme for \mol{226}{RaF} and simulation of corresponding one-dimensional molasses forces.
(a) Fortrat diagram indicating the ground state quantum number pairs $(N,J)$ participating in the possible transitions, where the four relevant transition branches are labeled in analogy to previous work in other molecules~\cite{Rockenhauser2023}.
	(b) Simulated Doppler-broadened spectrum (black line) of the transition \transitionXA{\nu=0}{\nu'=0} for a temperature of \SI{4}{\kelvin} with natural linewidth broadened transitions (blue line) using best-known experimental and \emph{ab initio} molecular constants summarized in the appendix and \cite{GithubLangenGroup}. 
	(c) Zoomed-in spectrum of the gray-shaded region in (b) to highlight the four hyperfine transitions (gray line and inset) that are used to drive the usual rotationally-closed cooling transition.
	(d)~Steady-state cooling force calculated using the OBEs as a function of velocity for a common one-dimensional molasses configuration with four laser sidebands addressing the respective hyperfine transitions in (c). The two combinations of magnetic field $B$ and total intensity $I$ shows the typical behavior of blue-detuned type-II level systems---large Doppler heating competing with small sub-Doppler forces for large magnetic fields and small intensities, whereas Sisyphus forces dominate over Doppler forces for smaller fields and increased intensities.
	}
	\label{fig:RaF_spectra}
\end{figure}

\subsection{Branching ratios and cooling forces for RaF molecules}

Recently,  radioactive molecular species have been created and investigated spectroscopically~\cite{GarciaRuiz2020,Udrescu2021,Udrescu2023}. Due to their deformed nuclei, heavy radioactive molecules offer high sensitivity for various tests of fundamental symmetries~\cite{Isaev2010,ArrowsmithKron2023}. Radium monofluoride is of high interest for such studies because it exhibits diagonal Franck-Condon factors and should thus be laser-coolable~\cite{Isaev2010,GarciaRuiz2020}. The odd isotopologues \mol{223}{RaF} and \mol{225}{RaF} are generally considered particularly promising, due to their high sensitivity to nuclear $P$- and $PT$-violation effects. 

We use the calculation of the eigenstates for these odd isotopologues as an example for the spectra calculation module included in \Py{MoleCool}. Whereas \mol{223}{RaF} features a more complex hyperfine structure due to the larger nuclear spin \( I_\text{Ra}^{223} = 3/2 \), \mol{225}{RaF}, with the smaller nuclear spin \( I_\text{Ra}^{225} = 1/2 \), exhibits a much stronger hyperfine coupling—on the order of several~\si{\giga\hertz}—approximately eight times larger than that of the former, based on the best currently available \emph{ab initio} calculations~\cite{Petrov2020,Skripnikov2020,stoneTABnuclearMoments}. Using in addition isotope-scaled constants from the well-characterized \mol{226}{RaF}~\cite{GarciaRuiz2020}, this hyperfine interaction leads to overlap of neighboring rotational states. Importantly, we find that the nuclear spins lead to a strong mixing of the $J'=1/2^+$ and $J'=3/2^+$ levels in the excited state \exs. This results in detrimental branching losses into higher rotational ground states, such as \gs$(N=3)$, for which selection rules no longer ensure rotational closure. Such losses from the rotationally closed cycle typically occur after only $\sim $ 13 scattered photons in \mol{225}{RaF} and $\sim $ 100 in \mol{223}{RaF}, respectively. The observed behavior is similar to the situation discussed for YbOH above and makes laser cooling highly challenging for both species~\cite{Pilgram2021,Augenbraun2023,Zeng2023}.

We note that this conclusion is highly sensitive to the values of the molecular constants used in our calculation. Additional high-resolution spectroscopy is therefore essential to evaluate other strategies for laser cooling of the odd isotopologues of RaF. We also point out that the behavior seen in RaF and YbOH is not generic: Certain structurally similar isotopologues of other heavy molecules, such as ${}^{137}$BaF~\cite{Kogel2021}, exhibit smaller excited-state hyperfine mixing, enabling the scattering of thousand of photons—sufficient for efficient laser cooling~\cite{Kogel2025}.

Due to the challenges associated with the level structure in the odd isotopologues of RaF, in the following, we restrict ourselves to its even isotopologues for the study of laser cooling in these molecules. In \mol{226}{RaF}, the nuclear spin $I^{226}_\text{Ra}=0$ vanishes and mixing losses are strongly suppressed. Based on high-precision spectroscopic results and these favorable properties, a basic laser cooling scheme for \mol{226}{RaF} has recently been proposed~\cite{Udrescu2023}. Here, we complete this scheme and model the laser cooling forces for \mol{226}{RaF}, incorporating the small but important hyperfine structure arising from the fluorine nuclear spin ($I_F = 1/2$). Setting up the calculations for this isotopologue also serves as an example for how to set up simulations for a new species from scratch in \Py{MoleCool}. Since the corresponding hyperfine structure constants have not yet all been experimentally measured, we use parameters obtained through high-precision \emph{ab initio} calculations~\cite{Petrov2020} to calculate the branching ratios of the individual levels~\cite{Kogel2021}. A summary of the constants employed is given in the appendix. 

The resulting Doppler-broadened spectrum of \mol{226}{RaF} at a temperature of \SI{4}{\kelvin} is depicted in Fig.~\ref{fig:RaF_spectra}a,b. Notably, the $\Lambda$-doubling in the \exs\ state exceeds the rotational splitting, resulting in the spectral positions of the P$_1$ and $^\text{P}$Q$_{12}$ branches~\footnote{For details on the labeling convention see, e.g. Ref.~\cite{Rockenhauser2023}} being interchanged with those of the Q$_1$ and $^\text{Q}$R$_{12}$ branches, as highlighted in the Fortrat diagram in Fig.~\ref{fig:RaF_spectra}a. This leads to an overlap of the usual laser cooling transition \gs($N=1$) $\rightarrow$ \exs($J=1/2^+$) with other transitions connected to higher rotational levels $N>1$ of the ground state.

For successful laser cooling, sustained optical cycling is ensured when no additional transitions are driven that transfer population from the $N = 1$ ground state to other rotational states ($N \ne 1$) via excited states with $J' \ne 1/2^+$. The only potentially disruptive transitions—Q$_1(N=1)$ and $^\text{Q}$R$_{12}(N=1)$—which couple directly to the $J' = 3/2^+$ state-are detuned by at least \SI{990}{\mega\hertz}, and therefore only lead to a negligible number of unwanted excitations. Moreover, when other rotational states $N\ne 1$ are off-resonantly addressed through overlapping transitions via higher excited states $J'>1/2$, their populations are depopulated after only a few photon scattering events, without affecting the optical cycle. The situation is equally favorable for the first repumping and depumping transition \transitionXA{\nu=1}{\nu'=0} and \transitionXA{\nu=0}{\nu'=1}, respectively, where the latter can be used for background-free detection~\cite{Rockenhauser2023}.

In contrast to the rotational structure, the molecular hyperfine structure in this isotopologue is relatively simple and exhibits significantly larger energy-level splittings compared to those of more typical diatomics such as CaF, SrF, or BaF. This results in a straightforward selection of frequency sidebands for laser cooling, as the individual sidebands are well-detuned from undesired transitions and thus operate with minimal interference~\cite{Kogel2025optimization}.

For modeling laser cooling forces, we can thus restrict ourselves to the four ground state hyperfine levels with $F=2,1^+,0,1^-$ of the cycling transition \cooltrans, as illustrated in Fig.~\ref{fig:RaF_spectra}c. The corresponding excited state levels $F'=0,1$ are unresolved. Following a common approach in molecular laser cooling experiments~\cite{Lim2018,Barry2014,Tarbutt2015a,Devlin2018}, each of these transitions is addressed using an individual laser sideband, which all share a common detuning $\Delta$. The steady-state cooling forces are then computed for a one-dimensional optical molasses setup, where four equally strong sidebands with a blue detuning of $\Delta=+2\,\Gamma$ are applied.
These calculations are performed in a large parameter space, involving transverse velocities $v$, total intensities $I$, and magnetic field strengths $B$ and laser polarizations, to identify suitable values that yield large laser cooling forces. 

Figure \ref{fig:RaF_spectra}d presents two examples of the resulting velocity-dependent cooling force, illustrating another characteristic behavior of multi-level type-II systems~\cite{Devlin2018,Kogel2021}. At higher magnetic fields, the Doppler force dominates over a broad velocity range and remains effective even at low laser intensities. In contrast, the magnetically-assisted laser cooling force is most efficient at lower magnetic fields, where it destabilizes dark magnetic sublevels. Its strength increases with laser intensity, provided the remixing rate is sufficiently fast. However, Sisyphus cooling competes with Doppler heating for blue detuned light in type-II systems and vise versa for red detuning. Consequently, it is possible either to apply a Doppler cooling force to capture a large amount of molecules with a broad range of transversal velocities or to cool fewer molecules to sub-Doppler temperatures.

Notably, the maximum achievable force, which is related to the scattering rate discussed in Sec.~\ref{sec:multileveldynamics} and given for the present $12+4$ level system by $F_\text{max} = 0.5\,\hbar k \Gamma/2$, is not attained here. This is because individual sideband frequency components, despite the large detunings discussed, remain effectively simultaneously blue- and red-detuned with respect to specific transitions, which limits the overall magnitude of the force.

\begin{figure}[tb]
	\centering
    \includegraphics{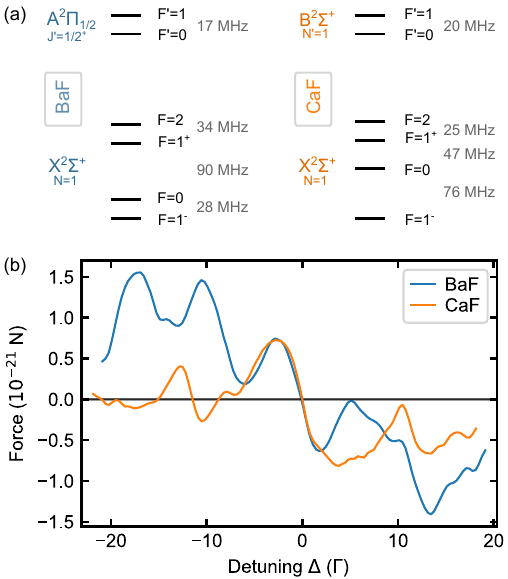}
	\caption{(a) Comparison of the hyperfine structure spacings in BaF (blue) and CaF (orange) for typical laser cooling transitions. The 12+4 level structure shown is identical to the one introduced in Fig.~\ref{fig:BaF_Ne_Bfield-I-3angles}a, but takes into account the individual level spacings of the species. (b) Corresponding laser cooling forces in BaF  and CaF molecules obtained from the OBEs. Detuning in this figure is defined relative to the zero-crossing of the force and is normalized to the respective linewidths $\Gamma$. While the forces in BaF primarily show the usual dispersion-like behavior, with a linear central region and falling-off wings, in CaF several of these dispersion features are visible. These arise whenever the detuning becomes comparable to a particular splitting of the hyperfine structure, which always depends on the specific structure of the molecule and is very pronounced in CaF. Powers in the simulations were chosen significantly higher in BaF ($I=1700\,I_\text{s}$) than in CaF ($I=100\,I_\text{s}$) to create forces of comparable magnitude.}
    \label{fig:fig7_bafcaftransversal}
\end{figure}

\subsection{Comparing transverse cooling for\\BaF and CaF molecules}
For a more detailed illustration of the consequences of such simultaneous blue and red detuning, and the subtleties of specific hyperfine splittings, we compare transverse laser cooling of BaF and CaF molecules. These species feature very similar structures, differing only slightly in their respective hyperfine structure splittings of the \gs\ ground states, as shown in Fig.~\ref{fig:fig7_bafcaftransversal}a. 

However, these subtle differences can lead to distinctively different force profiles. To highlight this, we simulate the cooling forces for the \gs $\rightarrow$ \exsb\ transition in CaF, as previously studied in~\cite{Rich2023}, as well as \gs $\rightarrow$ \exs\ in BaF, as studied, e.g., in~\cite{Rockenhaeuser2024lasercooling,Kogel2025optimization}. In both cases, we use simple equidistant sideband spectra, as can be  generated using a single EOM. Modulation frequencies of $\approx 11\,\Gamma$ and $\approx 14\,\Gamma$ were used for BaF and CaF, respectively.

The results are shown in Fig.~\ref{fig:fig7_bafcaftransversal}b. While BaF predominantly exhibits the expected behavior of a central, dispersion-like zero crossing accompanied by broad cooling and heating wings, CaF displays a much richer structure with an additional second cooling feature. Such additional features give rise to cooling forces at finite velocities. This can cause, for example, complex loading dynamics for magneto-optical traps, when loaded from transversely laser-cooled molecular beams~\cite{Rich2023}. 

The origin of these features lies in detuned sidebands that become resonant with different neighboring transitions, especially when the detuning matches the sidebands' modulation frequency. A similar effect can also be observed in BaF, where distinct peaks appear superimposed on the broad cooling and heating features. However, unlike in CaF, the specific energy splittings in BaF ensure that contributions from different transitions always add constructively for the given specific sideband configuration, thereby preventing the emergence of additional zero crossings.

A detailed understanding of these mechanisms is particularly crucial for efficiently combining slowing and transverse cooling in CaF, as this could boost the size of MOTs, increase achievable phase-space densities and enable Bose–Einstein condensation of these dipolar molecules by direct laser cooling. 
We note that strategies for efficiently combining slowing and transverse cooling have also recently been discussed in detail for SrF~\cite{Langin2023}.

\subsection{Bichromatic forces for\\BaF and CaF molecules}

As a final example, we compare a different type of force—the stimulated \emph{bichromatic force} (BCF)~\cite{Metcalf2017}—for BaF and CaF molecules. As discussed above, the relatively large mass of BaF together with its limited scattering force leads to only modest accelerations and correspondingly long slowing distances for typical molecular beams, highlighting the need for alternative methods such as the BCF.

So far, experimental demonstrations of bichromatic beam slowing have only been made for atomic species like cesium~\cite{Soding1997} and helium~\cite{Partlow2004,Chieda2012} but several theoretical simulations and proposals for molecular slowing have been put forward~\cite{Chieda2011,Galica2013,Aldridge2016,Yang2017,Yin2018,Wenz2020a}. The high effectiveness of BCFs for molecules has been demonstrated in transverse deflection experiments using strontium monohydroxide (SrOH)~\cite{Kozyryev2018a} and CaF molecules~\cite{Galica2018a}, motivating our comparison of the situation in BaF and CaF. We note that simulations of the forces for BaF have also previously been performed in the context of precision measurements using matrix-implanted molecules~\cite{Marsman2023}.

\begin{figure}[tb]
	\centering
	\includegraphics[width=0.45\textwidth]{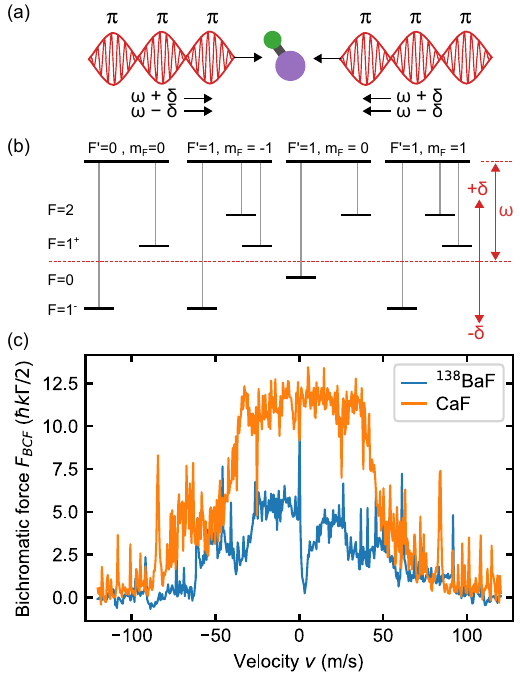}
	\caption{(a) Simplified representation of the BCF as a train of $\pi$-pulses acting on a molecule. The $\pi$-pulses are created using a beat pattern formed by the interference of two counter-propagating, bichromatic laser beams, each containing frequency components detuned by $\pm\delta$ around a common frequency $\omega$. (b) Four subsystems defined by isolating $\Delta m_F=0$ transitions on the cooling transitions of BaF and CaF molecules~\cite{Aldridge2016}. Energy splittings are not to scale. The symmetric detuning $\delta$ of the two laser frequencies used to create the BCF is indicated relative to the center of mass of the transition energies. (c) Comparison of bichromatic forces in BaF (blue) and CaF (orange) at a bichromatic detuning of $\delta = 75\Gamma$. In both species, the maximum forces obtained at a bichromatic phase of $\chi=\pi/2$ significantly exceed those in regular laser cooling and are active over a wide velocity range. CaF exhibits both a stronger force and a wider velocity range, whereas BaF shows an unusual resonance at small positive velocities, leading to a pronounced asymmetry in the force profile. This feature highlights how  the BCF arises from a complex interplay of species-specific level structure.}
	\label{fig:bichromatic}
\end{figure}

BCFs are created using a beat pattern of two dual-frequency laser beams. These beams are engineered so that the molecules effectively experience a constant train of $\pi$-pulses (see Fig.~\ref{fig:bichromatic}a). By choosing an appropriate relative phase $\chi$ between these pulses, this pulse train leads to a periodic sequence of excitation and stimulated deexcitation processes. Due to the stimulated nature of deexcitation by the $\pi$-pulses, the magnitude of the force is no longer limited by the excited-state decay rate. In addition, branching losses into unwanted ground states are strongly suppressed. In an idealized two-level system, the resulting force is given by $F=\hbar k \delta/\pi$, where $\delta$ is the symmetric detuning of the two laser frequencies from the transition frequency $\omega$. For sufficiently large detunings, the magnitude of this force can thus be significantly larger than the magnitude of the conventional scattering force, which is limited to $F=\hbar k R_\text{sc}$ (see Eq.~\ref{eq:R_sceff}).

As discussed previously, CaF and BaF share similar rotational and hyperfine structures, suggesting that similar BCF parameters might initially be expected to be applicable in both species. However, as we explain below, several general features of the BCF must be taken into account to ensure efficient implementation in generic multi-level systems and in BaF in particular.

More specifically, in multi-level systems with several ground and excited states and different branching ratios, the Rabi frequencies $\Omega$ (see Eq.~\ref{eq:Rabifreq_lup}) associated with individual magnetic hyperfine transitions generally differ. As a result, the $\pi$-pulse condition cannot be satisfied simultaneously for all transitions. In addition, because the hyperfine energies are not degenerate, the center frequency $\omega$ of the BCF beams cannot be tuned to resonance with every transition at once. One must therefore identify suitable compromises that optimally balance the contributions of the many transitions involved.

To achieve this, we observe that for a $2+1$ $\lambda$-type system with two nondegenerate ground states, BCFs are typically maximized in the regime of large detuning, where the detuning far exceeds the ground-state splitting~\cite{Aldridge2016,Aldridge2016PhD}. In this limit, the ideal Rabi frequency is given by the root-mean-square of the two individual Rabi frequencies.

In the multi-level systems of CaF and BaF, four approximately independent $\lambda$-type subsystems of this type can be isolated, each consisting of one excited magnetic hyperfine sublevel and the respective ground magnetic hyperfine levels, when driven with a linearly polarized light field (see Fig.~\ref{fig:bichromatic}b). For these subsystems, the quadrature sum of the individual elements of the electric dipole matrix is always $\kappa = 1/\sqrt{3}$ for both CaF and BaF.
This is due to normalization and symmetry in the angular momentum couplings. Consequently, all subsystems are expected to be driven efficiently with the same total Rabi frequency $\Omega_\text{tot} = \kappa \Gamma \sqrt{I/2I_\text{s}}$ (see Eq.~\ref{eq:Rabifreq_lup}).
This optimal Rabi frequency is obtained at a given detuning $\delta$ when the intensity of each bichromatic light-field component is set to $I = 3 I_\text{s} (\delta / \kappa \Gamma)^2$. 

For BaF, the ideal intensity $I$ at a given $\delta/\Gamma$ is about a factor of 10 lower than the corresponding value for CaF. At the same time, since the hyperfine manifold spans $\num{53.5}\,\Gamma$ in BaF compared to only $\num{23.2}\,\Gamma$ in CaF, maintaining the large-detuning regime generally requires choosing a larger detuning for BaF than for CaF. 

Going beyond this simplified picture, magnetic remixing of dark hyperfine states invalidates the assumption of four independent subsystems. As a result, the overall force magnitude is reduced from the idealized value, and the estimated parameters should only be regarded as a first approximation that needs to be further refined in full simulations~\cite{Aldridge2016,Aldridge2016PhD}.

Taking all these considerations into account, we simulate the BCF using OBEs, as rate equations cannot describe this coherence-driven mechanism. To reduce computational complexity and reach steady-state dynamics, we restrict the model discussed in this example to the ground vibrational states ($\nu=\nu'=0$) without including loss channels. We use the \gs $\rightarrow$ \exs\ for BaF and \gs $\rightarrow$ \exsb\ transition for CaF, resulting in the same 12+4 systems as in the previous example. To be in the large detuning regime for both species, a bichromatic detuning of $\delta=75\Gamma$ is applied, with laser intensities set to satisfy the $\pi$-pulse condition for BaF ($29.6\,$W/$\textrm{cm}^2$) and CaF ($280\,$W/$\textrm{cm}^2$). The bichromatic phase is set to $\chi=\pi/2$ to obtain the maximum force magnitude. Additional parameters are a magnetic field $B=\SI{20}{\gauss}$ and a polarization angle of $\theta=\SI{60}{\degree}$. 

The results in Fig.~\ref{fig:bichromatic}c show that both species experience strong forces over a broad velocity range, with those for CaF exceeding those for BaF by roughly a factor of two in magnitude. Interestingly, BaF exhibits an unusual destructive resonance at small positive velocities, which strongly breaks the velocity symmetry. This phenomenon arises from the intrinsic interplay between BaF's hyperfine structure and individual transition strengths. We also observe additional velocity-dependent resonances at fractional Doppler shifts of the bichromatic detuning ($v = {1/5, 1/4, 1/3,\dots}\times \delta/k$), with $k$ the wavevector of the bichromatic laser beams. 

The mean excited-state population is \SI{12.7}{\percent} in CaF but significantly higher (\SI{23}{\percent}) in BaF. The latter value approaches the theoretical maximum of $\Nu \approx \SI{25}{\percent}$ for a $12+4$ system. Such an increased excited-state population can lead to enhanced spontaneous vibrational branching losses, exceeding the simple $\pi$-pulse picture of the BCF. In practice, these losses would need to be addressed with a vibrational repumper, which in turn reduces the achievable BCF magnitude by a factor that depends on the precise branching ratios~\cite{Aldridge2016,Aldridge2016PhD}. However, this reduction can be circumvented by repumping through a separate electronic excited state.

Both findings, BaF-specific resonance and significantly different excited-state populations, demonstrate that BCFs strongly depend on details of the molecular structure. Consequently, BCF-behavior must be assessed on a species-by-species basis rather than inferred from simplified scaling arguments.

Finally, we apply our results to molecular beam slowing of BaF from an initial forward velocity of $v= 190\,$m/s~\cite{Albrecht2020} to rest. This initial velocity is much higher than the range of the calculated force profile. The laser frequencies must therefore be frequency-chirped to shift the force profile along the velocity axis as the molecules are slowed down, similar to the strategy in radiation pressure frequency-chirped slowing. In this process, the steep slope around $v\approx 0\,$m/s together with the almost vanishing force for $1\,$m/s $ < v < 10\,$m/s can be exploited to generate a highly compressed velocity distribution with a well-defined mean velocity during slowing. 

Effectively, the slowing is dominated by BCFs in the velocity range $-25~\text{m/s} < v < 0~\text{m/s}$, where the force reaches its maximum value of $F_\text{BCF} \approx 5 \hbar k \Gamma / 2$ for BaF. This results in a slowing distance of $d = \SI{14}{\centi\metre}$, which is about 20 times shorter than for radiation-pressure–based frequency-chirped slowing, and thus promises substantially higher numbers of slowed molecules.

\section{Conclusion}

In summary, we have introduced \Py{MoleCool}, a flexible and efficient toolbox for the modeling of light–matter interactions in high complexity atomic and molecular systems. By combining rate-equation and optical Bloch–equation approaches, the package enables comprehensive studies ranging from textbook two-level problems to realistic cooling schemes involving hundreds of coupled states,  demonstrating the relevance of \Py{MoleCool} for both educational purposes and advanced research. 

Going beyond these examples, the package is ideally suited for detailed optimization of forces and scattering rates, for example, using Bayesian optimization with Python’s \texttt{scikit-optimize} package—allowing users to efficiently identify optimal parameters for cooling~\cite{Kogel2025optimization, Rockenhaeuser2024lasercooling}. Furthermore, it could also be applied to the modeling of complex cooling schemes for exotic atoms~\cite{Zheng2025}.  

Due to its modular structure, the package also supports modeling decays through intermediate levels, which is important to describe, for example, leakage through intermediate $A'\Delta$ states in yttrium monoxide and BaF molecules~\cite{Yeo2015,Albrecht2020}. This is, for example, important for the simulation and optimization of magneto-optical traps of these species~\cite{Collopy2018,Devlin2018}, which is readily possible with our toolbox~\cite{Langlin2023}. Realistic  temperature estimates can be included using an approach based on the Fokker--Planck equation, which uses both the calculated force profiles and the excited-state populations as input~\cite{Devlin2018,Kogel2021}. 

Beyond molecules, this possibility to include multiple excited states also enables the detailed modeling of Rydberg excitation schemes commonly used for nonlinear quantum optics~\cite{Pritchard2013}, many-body physics and quantum computation~\cite{Browaeys2020,Cong2022} or coherent dynamics in hot atomic vapours~\cite{Huber2011}. 

Looking ahead, \Py{MoleCool} thus provides a foundation for further extensions, supporting the continued development of laser cooling in a variety of complex systems.

\section*{Acknowledgments}
We are indebted to Tilman Pfau for generous support and Eric Norrgard, Max M\"ausezahl and Einius Pultinevicius for fruitful discussions. This project has received funding from the European Research Council (ERC) under the European Union’s Horizon 2020 research and innovation programme (Grant agreements No. 949431 and 101113334), the RiSC programme of the Ministry of Science, Research and Arts Baden-W\"urttemberg and Carl Zeiss Foundation. It was funded in whole or in part by the Austrian Science Fund (FWF) 10.55776/PAT8306623 and 10.55776/COE1. T.G. acknowledges support by the Austrian Academy of Sciences (\"OAW) through a DOC Fellowship.
\section*{Appendix}
\subsection{RaF constants}
Below we summarize the molecular constants for the ground \gs\ and first excited electronic state \exs\ of \mol{226}{RaF}, as used in this work. The electronic offset is calculated via 
\begin{equation}
    T'_e \approx T_{00} - \omega'_e/2 + \omega_e + A'_{00}/2,
\end{equation} where primed constants belong to the excited state. To obtain a similar set of constants for the odd isotopologues \mol{225}{RaF} and \mol{223}{RaF}, isotope scalings are applied and combined with hyperfine structure constants computed in \emph{ab initio} calculations~\cite{Skripnikov2020,Petrov2020,stoneTABnuclearMoments}.	The upper limit of the lifetime $\tau\le\SI{50}{\nano\second}$ yields a natural linewidth of $\Gamma\ge 2\pi\times\SI{3.183}{\mega\hertz}$~\cite{GarciaRuiz2020}. A sample calculation to derive branching ratios and hyperfine losses for \mol{225}{RaF} and \mol{223}{RaF} is included in the repository of our package~\cite{GithubLangenGroup}.\\~\\

\begin{table}[h]
	\small
	\centering
	\begin{threeparttable}
		\caption{Constants for \mol{226}{RaF} used in this work.}
		\begin{tabular}{>{\centering\arraybackslash}p{0.4\columnwidth} >{\centering\arraybackslash}p{0.4\columnwidth}}
			\toprule
			\multicolumn{1}{l}{$\mathrm{X}^2\Sigma^+$}\\ 
			$\omega_e$                & 441.8000\tnote{a}  \\
			$\omega_e \chi_e$         & 1.6711\tnote{a}    \\
			$10^3\, B_e$              & 192.5175\tnote{b}  \\
			$10^7\, D_e$              & 1.4000\tnote{a}    \\
			$10^3\, \alpha_e$         & 1.0650\tnote{b}    \\
			$10^3\, \gamma$           & 5.8500\tnote{b}    \\
			$10^3\, b_F$ (F)          & 3.2133\tnote{c}    \\
			$10^3\, c$ (F)            & 0.6338\tnote{c}    \\
			\addlinespace 
			\multicolumn{1}{l}{$\mathrm{A}^2\Pi$}\\ 
			$T_e$                     & 14321.3770\tnote{b}\\
			$\omega_e$                & 435.5000\tnote{a}  \\
			$\omega_e \chi_e$         & 1.6350\tnote{a}    \\
			$10^3\, B_e$              & 191.5375\tnote{b}  \\
			$10^7\, D_e$              & 1.4000\tnote{a}    \\
			$10^3\, \alpha_e$         & 1.0450\tnote{b}    \\
			$A_e$                     & 2067.6000\tnote{a} \\
			$p+2q$                    & -0.4107\tnote{b}   \\
			\bottomrule
		\end{tabular}
		\label{tab:constantsRaF}
		\begin{tablenotes}
			\item[a] Reference~\cite{GarciaRuiz2020} using the relations $\omega_e \chi_e\approx\omega_e^2/(4D_e)$ and $A_e\approx A_{00}$.
			\item[b] Reference~\cite{Udrescu2023} using the relations $B_e\approx 3/2\,B_0 - 1/2\, B_1$ and $\alpha_e \approx B_0-B_1$.
			\item[c] Reference~\cite{Petrov2020} where the hyperfine tensor components are converted into Frosch and Foley parameters.
		\end{tablenotes}
	\end{threeparttable}
\end{table}

\bibliography{biblio}

\end{document}

%% file: CodeStructure.tex
%
\usetikzlibrary{positioning,shapes,shadows,arrows,arrows.meta,shapes.geometric,shapes.misc}

	\tikzstyle{abstract}=[rectangle, draw=black, rounded corners, fill=blue!30, drop shadow,
	font=\sffamily, anchor=north, text=black, text width=5cm, rectangle split parts=3, align=flush left]
	\tikzstyle{comment}=[rectangle, draw=black, rounded corners, fill=yellow!20, drop shadow,
	text centered, anchor=north, text=black, text width=2cm]
	\tikzstyle{arrowComp}=[{Straight Barb}-{diamond},>=triangle 90, thick ]
	\tikzstyle{arrow}=[-{Straight Barb},>=triangle 90, thick ]
	\tikzstyle{arrowInh}=[->,>=open triangle 90, thick ]
	\tikzstyle{textInh} = {near end, font=\ttfamily}
	\tikzstyle{line}=[-, thick]
	\begin{center}
		\begin{tikzpicture}[node distance=2cm,auto]
			\node (System) [abstract, rectangle split, rectangle split parts=3]
			{\hyphenchar\font=-1
				\large{System}
				\nodepart{second}
				calc\_OBEs(),
				calc\_rateeqs(),
				calc\_trajectory(),
				calc\_Rabi\_freqs(),
				plot\_all(), plot\_N(), plot\_Nsum(), plot\_Nscatt(), plot\_Nscattrate(), plot\_dt(), plot\_v(), plot\_r(), plot\_F(), draw\_levels(), plot\_spectrum()
				\nodepart{third}
				lasers, levels, Bfield,
				Nscatt, Nscattrate, photons, F, N, N0, t, v, r, v0, r0,
				steadystate, multiprocessing
			};
			\node (Levelsystem) [abstract, rectangle split,  below =1cm of System]
			{\hyphenchar\font=-1
				\large{Levelsystem}
				\nodepart{second}
				add\_add\_levels(), add\_electronicstate(),
				calc\_branratios(),
				calc\_dMat(),
				calc\_freq(),
				calc\_muMat(),		
				calc\_M\_indices(),	calc\_Gamma(),
				print\_properties(), reset\_properties(), plot\_transition\_spectrum()
				\nodepart{third}
				states, electronic\_states, grstates, exstates,
				dMat, dMat\_red, wavelengths, vibrbranch,
				N, lNum, uNum, load\_constants, Isat, Isat\_eff
				
			};

			\node (Bfield) [abstract, rectangle split,  left=1cm of System, yshift=1cm]
			{\hyphenchar\font=-1
				\large{Bfield}
				\nodepart{second}
				turnon(), turnon\_earth(),
				reset(),
				get\_remix\_matrix()
				\nodepart{third}
				strength, direction, angle, axisforangle, Bvec\_sphbasis,
			};

			\node (Lasersystem) [abstract, rectangle split,  below=1cm of Bfield]
			{\hyphenchar\font=-1
				\large{Lasersystem}
				\nodepart{second}
				add(), add\_sidebands(), get\_intensity\_func(), I\_tot(),
				getarr(), plot\_I\_1D(), plot\_I\_2D(), check\_config(), plot\_spectrum()
				\nodepart{third}
				pNum, I\_sum, P\_sum, freq\_pol\_switch
				
			};
			\node (Laser) [abstract, rectangle split,  below=1cm of Lasersystem]
			{\hyphenchar\font=-1
				\large{Laser}
				\nodepart{third}
				omega, I, P, k, kabs, lamb, f, E, phi, r\_k, w, FWHM, beta, f\_q, 		pol\_switching, pol, freq\_Rabi
			};

			\node (ElectronicGrState) [abstract, rectangle split,  right=1cm of Levelsystem, yshift=-1.3cm]
			{\hyphenchar\font=-1
				\large{ElectronicGrState}
				\nodepart{second}
				add\_lossstate(), del\_lossstate(), print\_remix\_matrix()
				\nodepart{third}
				has\_lossstate
			};

			\node (ElectronicExState) [abstract, rectangle split,  above=0.6cm of ElectronicGrState]
			{\hyphenchar\font=-1
				\large{ElectronicExState}
				\nodepart{second}
				\nodepart{third}
				Gamma
			};

		\node (ElectronicState) [abstract, rectangle split,  above=0.8cm of ElectronicExState]
		{\hyphenchar\font=-1
			\large{ElectronicState}
			\nodepart{second}
			add(), state\_exists(), load\_states(), draw\_levels(),
			print\_properties(), set\_init\_props()
			\nodepart{third}
			states, N, freq, gfac, v\_max
		};

			\node (State) [abstract, rectangle split,  above=1cm of ElectronicState]
			{\hyphenchar\font=-1
				\large{State}
				\nodepart{second}
				is\_equal\_without\_mF(), copy(), check\_QuNrvals()
				\nodepart{third}
				QuNrs, is\_lossstate, QuNrs\_without\_mF			
			};

			\draw[arrowComp] (Bfield) -- node [near end,above] {} (System);
			\draw[arrowComp] (Lasersystem) -- node [text width=0.3cm,near end,above] {} (System);
			\draw[arrowComp] (Laser.north) -- node [near end,font=\ttfamily] {} (Lasersystem.south);
		
			\draw[arrowComp] (Levelsystem) -- node [near end] {} (System);
			
			\draw[arrowComp] (State) -- node [near end,font=\ttfamily] {} (ElectronicState);
			\draw[arrowComp] (ElectronicExState.west) -- node [near end,font=\ttfamily] {} (Levelsystem);
			\draw[arrowComp] (ElectronicGrState.west) -- node [near end,xshift=0.2cm,font=\ttfamily, above] {} (Levelsystem);	
			
			\draw[arrowInh] (ElectronicState) -- (ElectronicExState);
			\draw[arrowInh] ([xshift=2cm]ElectronicState.south) -- ([xshift=2cm] ElectronicGrState.north);
			
		\end{tikzpicture}
	\end{center}
